\documentclass[sigconf]{acmart}
\usepackage{changes}
\usepackage{url}
\usepackage{csquotes}
\usepackage{enumitem}
\usepackage{balance}
\usepackage{graphicx}
\usepackage{todonotes}
\usepackage[utf8]{inputenc}
\copyrightyear{2020} 
\acmYear{2020} 
\setcopyright{acmlicensed}
\acmConference[AIES '20]{Proceedings of the 2020 AAAI/ACM Conference on AI, Ethics, and Society}{February 7--8, 2020}{New York, NY, USA}
\acmBooktitle{Proceedings of the 2020 AAAI/ACM Conference on AI, Ethics, and Society (AIES '20), February 7--8, 2020, New York, NY, USA}
\acmPrice{15.00}
\acmDOI{10.1145/3375627.3375874}
\acmISBN{978-1-4503-7110-0/20/02}

\begin{document}
 \fancyhead{}
%
\title{Ethics of Food Recommender Applications}
\author{Daniel Karpati}
\affiliation{%
  \institution{University of Luxembourg}
}
\email{daniel.karpati@uni.lu}

\author{Amro Najjar}
\affiliation{%
  \institution{University of Luxembourg}
}
\email{amro.najjar@uni.lu}

\author{Diego Agustin Ambrossio}
\orcid{0000-0002-3424-931X}
\affiliation{%
  \institution{University of Luxembourg}
}
\email{agustin.ambrossio@uni.lu}

\begin{abstract}
The recent unprecedented popularity of food recommender applications has raised several issues related to the ethical, societal and legal implications of relying on these applications. In this paper, in order to assess the relevant ethical issues, we rely on the emerging principles across the AI\&Ethics community and define them tailored context specifically. Considering the popular Food Recommender Systems (henceforth F-RS) in the European market cannot be regarded as personalised F-RS, we show how merely this lack of feature shifts the relevance of the focal ethical concerns. We identify the major challenges and propose a scheme for how explicit ethical agendas should be explained. We also argue how a multi-stakeholder approach is indispensable to ensure producing long-term benefits for all stakeholders.
After proposing eight ethical desiderata points for F-RS, we present a case-study and assess it based on our proposed desiderata points.  
\end{abstract}


\maketitle



\section{Introduction}
Using computers to assist users in determining what they eat is an ongoing research topic for more than half a century. For instance, since the 60's \cite{balintfy1964menu}, \emph{Menu Planning} computer programs have been devised to help the users optimise the costs of their food with later solutions in the 70's integrating user preferences \cite{lancaster1992history}. 

Recently, the widespread and ubiquitous use of smartphones, as well as the establishment of large online and crowdsourced databases aiming to index database of food products from around the world, have dramatically changed food recommender systems with the emergence of specialized applications aiming to provide assessments and recommendations during the shopping process. While these applications aim to empower the user's \enquote{Right to Know} and are a clear manifestation of the recent trend of \enquote{information activism}, recent works in the literature have started discussing their impact of users' behavior and their underlying ethical considerations \cite{hansson2017promoting}. The facts that these applications \emph{(i)}~are being adopted by millions of users make their recommendations impactful on both producers and consumers, \emph{(ii)}~unlike traditional consumerist magazines \cite{mallard2007performance}, these applications make the relationship between consumers and the market less abstract since they are accessible virtually all the time, and \emph{(iii)}~in order to operate real-time, and provide information, assessment and recommendation on hundreds of thousands of products, these applications tend to make a compromise between scientific logic (\emph{i.e.} the score and assessment attributed to food items are not always grounded on scientific truth), technological uncertainty (\emph{e.g.} the information of products in the databases might be erroneous or incomplete) and consumer concerns (\emph{e.g.} costs and product availability) \cite{soutjis2019new}. 

Based on recent debates regarding the ethics of recommender systems \cite{karimi2018news,milano2019recommender,kampik2019technology}, the guidelines proposed by private and international committees \cite{jobin2019global,hleg2019ethics}, and on the latest research on food recommender systems \cite{tran2018overview}, this article investigates the relevant ethical framework for such systems, proposes the ethical desiderata, argue for multi-stakeholder approach to food recommender systems, and point out future research directions.

The rest of this article as follows.  Section~\ref{sec:background} provides the background on Ethics and AI and the literature of recommender systems. Section~\ref{sec:Desiderata} presents the desiderata for non-personalised food recommender systems. Section \ref{sec:Yuka} presents a case-study on the Yuka food recommender application and Section~\ref{sec:Conclusion} concludes this paper.
\section{Background}
\label{sec:background}
\begin{figure}
    \centering
    \includegraphics[width=\columnwidth]{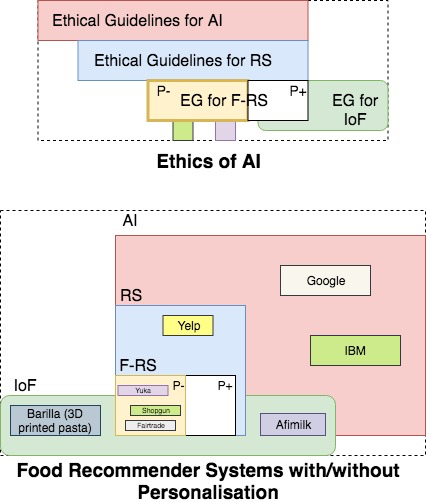}
    \caption{The different levels of ethical guidelines \& the matching AI context for food recommender systems}
    \label{fig:fig1}
\end{figure}
In this section, we are reviewing the literature by gradually moving from abstract principles to concrete case studies. In \ref{subsec:EthicsAI} we discuss our approach to generic ethical AI guidelines. In \ref{subsec:KeyConcepts} we identify the key principles emerging from these guidelines. By narrowing down our scope in \ref{subsec:EthicalRS} we identify the main concepts that will serve our starting frame for Section~\ref{sec:Desiderata}. Finally, Sections \ref{subsec:IoF} and \ref{subsec:CaseStudies} are pointing out the gap in the literature this article aims to cover.     
\subsection{Ethical AI Guidelines}
\label{subsec:EthicsAI}

Pushed the proliferation of AI applications and their use in different contexts and application domains, the last few years saw a tremendous increase in ethical guidelines for developing and implementing AI systems \cite{morley2019overview},\cite{taddeo2018ai}. Many big tech companies \cite{microsoft2017}, \cite{ourprinciples2018} and \cite{sony2018} as well as consulting firms \cite{pwc}, \cite{accenture2018} are publishing their \enquote{ethical guidelines} or \enquote{ethical principles}. Also national governments (e.g. UK \cite{houseoflords2018}, Singapore \cite{pdpc2018}), supranational entities \cite{hleg2019ethics} and international research institutes \cite{asilomar2017}, standards organisations~\cite{iec}, are also working to produce similar documents. 

In order to analyse these white papers and guidelines, several surveys and works in the literature aim to provide a broad overview of these various undertakings \cite{jobin2019global,hagendorff2019ethics}. Despite these efforts, the difficulty to evaluate and compare these guidelines lies mainly in the lack of clarity regarding the highly abstract and vague principles in these texts, which are in most cases not even close to the precision that would allow a more formal comparison and exploration of these architectures. Furthermore, this conceptual vagueness is not only present across the different guidelines, but it could also be found within the same proposed guideline\footnote{For instance, the ethical framework of the  EU HLEG expert group has contradicting takes on fairness: on the one hand, fairness is articulated as one of the four Principles (fairness as a Principle), but also it is part of one of the seven Requirements ensuring the Principles (fairness as a Requirement), and it is also noted within the same text that based on fairness we should be able to mitigate the trade-offs between not just the seven Requirements, but also between the Principles, one of which is fairness itself~\cite{hleg2019ethics}.}. While these broad ethical guidelines can be seen as an important first step towards a thoroughly considered practice for \emph{ex-ante} regulation, \emph{ex-post} adjudication and, hopefully, for the creation of actual AI technologies, still, the principles articulated in these guidelines should be situated in the actual context in order for them to be operationally applicable. (\emph{i.e.} bringing them 'down' closer to real-world applications and gradually clarifying their empirical meaning within more restricted domains. 
Similarly to our most fundamental rights in a given legal system or those are articulated in international declarations, such as the right to human dignity, we suggest approaching these broad concepts in the spirit that their meaning will become ever clearer as we 'zoom' into the actual context. As much as we are skeptical of the ability to design a technically understandable (computable) ethical framework for all AI-related applications 'deployable' anywhere regardless of the domain, and the geographical, societal and cultural circumstances, we are also an optimist that such ethical framework could \textit{evolve} through the cumulative works and reflections of the many stake-holders discussing their perspectives both horizontally and vertically. 

This working method in our case effectively means that we structure this section---after introducing the key concepts emerging from general ethical guidelines---to capture the relevant meanings of these principles considering recommender systems, then specifically explore the literature regarding food-recommender systems, and based on real-world applications available today, we further specify the context to provide (in section 3) the ethical desiderata for food-recommender applications that does not build on personally tailored recommendations (F-RS(P-)). Figure \ref{fig:fig1} depicts this method.

\subsection{The key concepts}
\label{subsec:KeyConcepts}
This paragraph introduces the key principles that are emerging from the many recently published guidelines. Intangible they are, similarly to the Universal Declaration of Human Rights, or GDPR; their merit lies in providing a broad overview that clarifies where further research and discussion are necessary to be pursued.

Jobin \emph{et al.} \shortcite{jobin2019global} by analysing 84 guidelines have identified eleven broad ethical desiderata. Among these, five principles were discussed in more than half of the papers subjected to their content analysis. Ordering by frequency, the eleven principles are the following: Transparency (73/84); Justice and Fairness (68/84); Non-Maleficence (60/84); Responsibility (60/84); Privacy (47/84); Beneficence (41/84); Freedom \& Autonomy (34/84); Trust (28/84); Sustainability (14/84); Dignity (13/84); Solidarity (6/84). It is important to note, that each of these concepts was understood broadly. For instance, the tag ‘Transparency’ included codes for ‘transparency', ‘explainability', ‘explicability', ‘understandability', ‘interpretability', ‘communication', ‘disclosure', and ‘showing. We would like to reinforce our commitment to a benevolent approach to understanding these high-level principles in their vagueness: we are not building a Gothic cathedral of abstract concepts but aim for a resilient, gradual understanding that can guide us ultimately on the ground level, \emph{i.e.} on the empirical battlefront of certain kind of AI applications and their environments. We kindly advise the reader to do the same, and focus their criticism there. We are also not going to place too much emphasis on the ranking of the principles simply because their relevance varies depending on the context.

\subsection{Ethical RS}
\label{subsec:EthicalRS}
In the literature \cite{burr2018analysis,karimi2018news,de2010identity}, Recommender systems (RS) are defined as systems that generally collect, organise and evaluate large quantities of (usually personal) data, and impact the users’ interaction with (individually) tailored experience of their digital realm.  Milano \emph{et al.} \shortcite{milano2019recommender} after reviewing the literature on RS articulated a taxonomy that enabled them to identify the key ethical issues usually emerging in RS literature. On the one hand, their taxonomy distinguishes between utility and rights-based concerns. While utility as a concept can usually be assessed quantitatively, concerns about rights violations are in most cases evaluated qualitatively. On the other hand, the taxonomy also separates concerns based on the temporality of the potentially harmful impact, distinguishing between immediate harm and the risk of potential future exposure. Based on their analysis of the literature, they identified six major concerns: (1) ethical content (\emph{i.e} how to implement certain values in RS); (2) privacy; (3) autonomy and personal identity; (4) opacity; (5) fairness; (6) polarisation and social manipulation. Using this double-faceted approach they classified (1) biased recommendations  (biased, in the sense of ethically biased based on the implemented ethical content) as an immediate negative impact on utility; (4) opacity and (1) questionable content creation as exposure to risk in RS; while (5) unfair recommendations and (3) the encroachment on individual autonomy and identity as immediate rights violation; finally (2) privacy and (6) social manipulability and polarisation as risk exposures. This framework (beyond its own merits) underscores the need and the benefits of a multi-stakeholder approach in RS in general. This approach is also recommended by other recent works in the literature \cite{abdollahpouri2017recommender,burke2016towards,liao2019building} since it systematically addresses issues of both imminent and future consequences, and it sheds some light on that ethical, societal and legal concerns related to RS are more complex than for us to be able to evaluate them only from the perspective of its immediate utility for the user.

\subsection{IoF, a different framework}
\label{subsec:IoF}
To the best of our knowledge, there is not such a comprehensive work on food recommender systems in the literature that undertakes the challenge to provide a contextual understanding of the generally prescribed ethical guidelines. The closest work to this date is Leone’s paper that identifies ethical challenges for smart systems in the agri-food domain \shortcite{leone2017beyond}. His contribution to the literature is to introduce an architecture of the Internet of Food (IoF) concept and outline the concerns regulators should aim to solve. Generally, IoF as a concept was envisioned as sophisticated communications and digital services between machines, consumers and companies through sensorisation to provide information about nearly all food ingredients and products \cite{brewer2019internet}. IoF ranges from monitoring and executing steps in the food production via the combination of advanced sensory applications, advanced methods for collecting and interpreting data, and automation to the creation of digital information platforms based on which AI systems can give personalised recommendations to the consumer \cite{leone2017beyond},\cite{boulos2015towards}. For this reason, IoF is not limited to food recommender systems and therefore the proposed ethical architecture hardly brings us any closer to our goal than the general ethical guidelines: with its design suggesting a continuous reinforcement of these criteria, the IoF architecture is visualised as a circle from Privacy that leads to Transparency making Education possible, which enables Negotiability that brings Agency which resolves Responsibility that leads back to Privacy \cite{leone2017beyond}. An ethical framework for evaluating food-RS still remains uncovered.

\subsection{Existing case-studies}
\label{subsec:CaseStudies}
Case-studies, however, emerged in the last two years examining certain applications(Fairtrade, Shopgun, Greenguide, Yuka), but they tend not to aim at formulating arguments on an abstract level; they do not create a framework for a \textit{type} of applications. Hansson's article explores three Swedish smartphone applications (Fairtrade app, Shopgun, and GreenGuide) that promote ‘ethical’ consumption \shortcite{hansson2017promoting}, more recently Soutjis wrote a detailed analysis of the fashionable Yuka application \shortcite{soutjis2019new}, that claims to recommend healthier products. These applications are made to help consumers to choose and buy products (in the case of Fair-trade app, Shopgun) that align with the ethical considerations or health conceptions (Yuka) of the makers, or to advise them to choose more environmentally-conscious practices (GreenGuide). Hence, in these cases, ethical apps are not meant to be understood as applications that were designed according to ethical guidelines but as applications that have a purposely in-built ethical (or other) agenda to shape consumer behaviour. The ethical considerations of the designers incorporated in these apps are based on an arbitrarily decided method (by the creators) resulting in an amalgam of expert advice and available product information (\emph{e.g.} qualification for certain labels and information obtained from the product’s barcode~\cite{hansson2017promoting, soutjis2019new}.

Although these applications are part of IoF as well as they are of RS, all of them lack the ability of personalised recommendations, which is a common and much-discussed feature of RS and most certainly part of the vision of IoF \cite{boulos2015towards}. Keeping that in mind, we would like to propose a framework that defines the ethical desiderata for these existing applications serving the designers and regulators of today, and laying the groundwork for further research.

\section{Ethical Desiderata for Food Recommender Systems without Personalisation}
\label{sec:Desiderata}
In this section, we make our case how further distinguishing between F-RS with personalisation (F-RS(P+)) and without (F-RS(P-)) alters the ethical focus. After reviewing the existing applications and finding that declaring the ethical desiderata for F-RS(P-) is the more pertinent task to adhere to, we climb two steps back on the 'abstraction ladder' and use the ethical framework developed for RS by Milano at al. as the reference point to outline the ethical desiderata for F-RS(P-).

\subsection{Existing Applications}
Generally speaking, the F-RS (food recommender systems) of today, conspicuously differ in one feature from other RS, namely, they don’t build on users' personal data or behaviours and recommend personally tailored solutions based on that. This is an important distinction and clearly demonstrates the alternate motivations behind these types of applications. The promise behind RS is that the user will be directed towards their interest, which is inferred from their actual customs provided in some form of data. The motivation behind F-RS of today (as has been demonstrated by the ones popular in the European market) is helping consumers to choose products based on some shared preferences mostly defined by their creators. In other words, the former has a universal value-neutrality: whatever you like, you will be able to find it, efficiently. The promise behind food recommender systems without personalisation (F-RS(P-)) is very different: there is a set of preferences regarding food consumption, if you share them, you will be able to find products satisfying the preferences, efficiently. The lack of personalised tailoring clearly shifts the focal ethical issues related to RS in the case of F-RS(P-), hence, F-RS(P-)  cannot be regarded just as a domain-restricted RS.
If we take a step back and look at the major concerns identified by Milano \emph{et al.} on a more abstract level for RS \cite{milano2019recommender} we can see how the lack of personalisation changes the relevance of those concerns, and also alter what those concerns mean in the specific context of F-RS(P-) empirically.

\subsection{Ethical Content}
In the case of RS, ethical content is a particular type of contents, we might rephrase it as ethically sensitive content. However someone tries to tackle this problem, the key issue is obviously that there are very different ethical convictions varying individually. Whether the proposed solution comes from applying filters \cite{bobadilla2013recommender} based on geographically located cultural norms \cite{souali2011automatic} or shifting the responsibility to the user and make them ‘set’ their preferences \cite{tang2016should}, the problem is addressed from the meta-level, before any specific set of ethical considerations are chosen. In the case of F-RS(P-), we are already committed to certain preferences, and this commitment to shared values with the designers grants the very existence of such an application.

\subsection{Privacy}
In relation to privacy concerns regarding F-RS(P-), the burden to balance a reasonable trade-off between accuracy and privacy, one of the key challenges in RS \cite{jeckmans2013privacy}, is nonexistent. ‘Accuracy’ as a concept when the recommendation does not change regardless of who the user is loses its connection to privacy, therefore the key issues regarding privacy today is focused on how data are collected, stored, shared and inferred from. These issues in the European context are already regulated with the adoption of the General Data Protection Regulations in May 2018 \cite{voigt2017eu}. 

\subsection{Autonomy and Personal Identity}
Similarly, without personalisation there is not algorithmic profiling and otherwise relevant arguments \cite{koene2015ethics} for reshaping the consumers' personal identity without them being aware of it (an issue extensively discussed in the recent literature on RS ethics c.f.  \cite{floridi2011construction}). However, certain arguments still need to be considered, depending on the business model of the particular application in question. Any food recommender system that generates its revenue through either by selling data or advertisements has the agenda to keep the user on the platform as long as possible. The method they use in the hope of realising this, however, should be made at least transparent, and in some cases, further regulative measures might have to be considered. We think the requirement of \enquote{informed consent} when entering a business relationship should entail the awareness about how the other party is compensated for the services they provide. Especially in the context of F-RS(P-) which explicitly build on shared ethical values, it seems an indispensable part to understand how the advertised ethical agenda and the compensation of the service provider can be pursued coherently. F-RS(P-) with an anti-consumption agenda, for instance, that get compensated through advertisements on their platform should inform their users on how these seemingly conflicting goals are resolved. 

\subsection{Opacity}
In this section, we propose a five-step scheme, a necessary requirement to avoid opacity, both regarding how the ethical agenda (Section~\ref{subsubsec:Opacity1}) and other optimising features (Section~\ref{subsubsec:Opacity2}) and their implementations should be explained.  

\subsubsection{Explaining the Ethical Agenda and its Implementation}
\label{subsubsec:Opacity1}
Regarding F-RS(P-) that promise to pursue an explicit ethical agenda, it is necessary to explicitly make the user aware of how such pursued values are understood and implemented. An ethical desideratum expressed in a natural language usually does not come with an unequivocal path of its implementation. Promoting the cheapest option for buying 1 kg of bread in half a mile radius is straightforward enough, but promoting the healthiest, the most environmentally friendly, the most fair-trade sensitive, the one that makes your skin beautiful, etc., are not so much. Engaging with an application that promotes such an agenda should rely on a shared commitment towards the empirically well-defined concept rather than simply committing to the phrase. To what extent the explanation should be provided is clearly one of the more complex questions regarding F-RS(P-). 
Regarding the explicit ethical agenda the explanation chain should answer these questions following this proposed scheme:

\begin{enumerate}
  \addtocounter{enumi}{-1}
    \item Phrase (Healthy, Green, Best Value, etc..)
    \item What is the criteria for a product to be considered as such?
    \item Who claims these criteria are necessary and sufficient?
    \item Based on what should they be trusted on this matter?
\end{enumerate}


The proposed \enquote{chain of explanations} regarding the ethical agenda and its implementation should serve to bridge the gap between \enquote{word of mouth} recommendations and those provided by F-RS. This often cited analogy \cite{herlocker2000explaining} has its shortcomings precisely for the lack of trust that is otherwise usually evaluated in the complex realm of human interactions. Another meaningful difference when considering this analogy is that even if, when \enquote{word of mouth} recommendations do not contain the interpersonal element, usually it is not provided by an agent that has built a system for the purpose of giving recommendations, which, even if not profit-oriented, most certainly needs financial resources to maintain operating. In this sense, we perceive the attitude towards F-RS(P-) would shift from the \enquote{word of mouth} analogy slightly towards advertisements.
Both arguments clearly show that we cannot dismiss the gap between F-RS(P-) and occasional \enquote{word of mouth} recommendations. 

\subsubsection{Explaining the Non-ethically Motivated Parts of the Algorithm}
\label{subsubsec:Opacity2}
Besides how the ethical agenda is implemented, we find also crucial to make the recommending algorithm explicitly understandable. This should happen at least by fulfilling two criteria:
\begin{enumerate}\addtocounter{enumi}{3}
    \item \label{enum:p1} Declaring all other objectives relevant to the recommendation by defining their purpose for the user and their consequences.
    \item \label{enum:p2} Declaring how these variables are affecting the process of recommendation, showing how the optimisation happens between the implemented ethical agenda and the further objectives.
\end{enumerate}


Therefore fulfilling the ethical desideratum to avoid opacity in F-RS(P-) are twofold: first, how the ethical agenda is understood and implemented following the scheme we proposed, and second, to explain the non-ethical part of the algorithm that affects the outcome as discussed in Points~(\ref{enum:p1}) and~(\ref{enum:p2}).

\subsection{Fairness}
\label{subsec:Fairness}
Research on AI fairness has recently received a strong momentum. These works range from calls for a fair and unbiased AI \cite{zou2018ai}, to other works aiming at understanding fairness in AI context \cite{verma2018fairness}, or trying to propose mechanisms aiming to operationalise it \cite{dwork2018decoupled}.
These works have been echoed in the recent literature of recommender systems \cite{abdollahpouri2019unfairness}. 
In the context of F-RS(P-), determining what concepts of fairness should be taken into account is far from obvious. We identified the following interpretations of fairness that we consider to hold relevance: 

\begin{enumerate}
    \item We propose that procedural fairness should be granted towards the manufacturers in two steps. First, by making accessible the information about the criteria upon which their product is evaluated and recommended. Note, that it should not come with any compromise given the rule-based nature of these algorithms and also considering that in these existing cases the mathematical/theoretical contribution is basically irrelevant. In other words, these are rather simple algorithms. Second, the designers should aim to complete and update the database they are working from. 
    
\item We propose that regarding the fairness of outcomes, F-RS(P-) should commit to proportional fairness based on the evaluation regarding their ethical agenda and whenever the ordering of the recommendations alter from the ordering of the evaluations, this should be made explicit for the user.

\item Regarding eliminating unwanted bias from F-RS(P-), we propose a dynamic multi-stakeholder environment, for F-RS(P-) the difficulty lies in determining what is considered as 'unwanted', and we think these decisions should be debated with relevant representatives of the general public being involved. Controversial our proposal might seem, yet given the potentially disruptive nature of F-RS(P-) to the food industry, which may have broad and unintended consequences related to our society, the involvement of such actors would benefit society in large.
\end{enumerate}

In the case F-RS(P-), we find it difficult to argue that any relevant and plausible notions of fairness regarding our societies can be maintained while benefiting of AI technologies without applying multi-stakeholder architectures. This statement, however, discussed in our next section.

\subsection{Polarisation and Social Manipulability}

In this section, we clarify the focal menaces for the society that occur with the prevalence of FRS(P-) and further argue for a multi-stakeholder approach to counter them. 
\subsubsection{Multi-stakeholder Approach}
\label{subsubsec:MultiStakeholder}

Contrary to RS in general, polarisation seems less of an issue, our concerns should, therefore, be focused more on manipulating the consumer by promoting an ethical agenda only by the name, but it is actually understood and implemented in a very oversimplified, arbitrary manner, which is somehow gets \enquote{lost in translation} due to the lack of proper explanations.  Unfortunately, the mere existence of these explanations is most probably not sufficient enough, for there is (or at least could be) a conflict between the long-term interests of the consumer, society and the creator of the application. This conflict could occur if we consider how important user experience in these F-RS(P-) to become commercially viable. Attracting as many users as possible behind an ethical agenda certainly seems easier if the stress is put on a catchy buzzword (\emph{e.g.} healthy, environmentally friendly, \emph{etc.}) rather than the interpretation and implementation of these values. For this reason, a multi-stakeholder approach to F-RS(P-) would seem necessary if we want to live up to our commitment to beneficial AI, for we surely want to avoid scenarios where it is essentially a race to the bottom on de facto providing, yet for the user presenting in the least accessible way the prescribed explanations. On the contrary, we need stakeholders who truly commit to their values and promote their interpretation and implementation. Hence, involving representatives of the general public, independent experts, that essentially have an interest in presenting these explanations in such a way that captures the attention of the user would seem beneficial to all stakeholders with serious, beneficial intent and would help integrate F-RS(P-) to the benefit for society.
\subsubsection{Disruptive effect on food industry}
\label{subsubsec:Disruptive}
 Another reasonable concern regarding social manipulations, that without ensuring the quality of F-RS(P-), the disruptive effect on our food sector can have severe consequences. Once, a particular application reaches certain popularity, it may have an effect of forcing producers to accommodate their products to the newly set standards. Obviously, without supervision, this could go either way. From a societal perspective, we need to make sure it enhances the quality of our lives, or raises certain ethical standards, and not distort it. The food industry and agriculture are fundamental pillars in our society that gets special attention from regulators, where a dynamic interaction between the state, producers, farmers, and consumers are established. If F-RS(P-) disrupt the status quo, we need to ensure, that it happens for the better.

\subsection{Robustness}
\label{subsec:Robustness}
Since F-RS(P-) are heavily built upon available data regarding the products, there is an added ethical concern to those identified on a higher level (RS) to comply with the principles expressed across the ethical guidelines. We have already discussed how crucial to aim for updated, complete, verified, validated data sets regarding the products in the given domain to fulfill procedural fairness. However, we also need to put emphasis on building robust data sets that cannot be easily manipulated.

\section{Case-study: Evaluating Yuka}
\label{sec:Yuka}
Yuka is a mobile application for food recommendations. Created as a startup in 2016, the application, according to its homepage, \enquote{scans your products and analyzes the impact on your health. In the blink of an eye, it deciphers labels for you: you see the products that are good and those that are best avoided} \cite{Yuka}.

In order to provide its assessment on the shopping products, Yuka relies on the Open Food Facts (OFF) \cite{OFFProducts}. The latter is an online and crowdsourced database of food products from around the world. Later, Yuka started also developing their own database.
The widespread use of the application by millions of customers in France and the neighbouring countries has raised several comments and criticism from the food industry, nutritionists, the media, and the research community \cite{soutjis2019new, vayre2019intelligence}: 
\begin{itemize}
    \item \textbf{Product assessment approach} The scoring system employed by Yuka relies on three elements: \emph{(i)} the score a product gets from Nutriscore \cite{chantal2017development}, \emph{(ii)} the presence of additives in the products, and \emph{(iii)} if the product is organic. Unlike Nutriscore which for nearly a decade was debated by expert groups and was validated by the relevant French legislative body, Yuka's methodology was never put up against such scrutiny. On additive its stance is controversial; since, for instance, many of the additives sanctioned by Yuka are not regarded as harmful by the relevant regulatory bodies of the EU, and there is no scientific evidence behind many of their claims. Similarly, Yuka automatically gives an extra 10\% score to any organic product despite the fact that the positive effects of organic products on the human health are still subject to scientific debate. Even more concerning the absolutely arbitrary manner they combine these factors: 60\% for Nutriscore, 30\% for the existence of certain additives and 10\% if the product is organic. According to our scheme, this should be made clear for the user (see points (2) and (3) in Section~\ref{subsubsec:Opacity1}), namely, who are the natural persons behind this calculation? What makes them reliable? Are they qualified to make such an assessment? 
    Everybody is entitled to their opinion, surely, but if the methodology of assessing 'healthy' products is based on the hunches of 'Grandma Gourmet', then it should be duly noted. 
    \item \textbf{Technical issues:} Yuka relies on crowdsourced databases whose main source of data is the nutrition facts written on the product packing. Yet, these databases are subject to erroneous and outdated input. In addition, food manufacturers are allowed to change the ingredients (with certain restrictions, for instance, adding allergens) of their products without necessarily changing the barcode of the product. Consequently, in some cases, the score given by Yuka can be either imprecise or outdated. For these reasons, the ethical desiderata we outlined in Section~\ref{subsec:Robustness} are not fulfilled, and certain elements considering fairness (see below) are also violated.  
    \item \textbf{Lack of explainability}: In addition the score it gives to products, Yuka also recommends an alternative product in the case a given product has fared below an overall score of 50. However, despite the growing interest in eXplainable AI and its impact on enhancing user trust and acceptability \cite{anjomshoae2019explainable}, as well as explainable and fair recommender systems \cite{zhang2018explainable, koutsopoulos2018efficient},  Yuka does not explain why a particular product is chosen as an alternative. Conspicuously, the recommended alternatives are not ranked by their score but other factors such as 'product availability' (which is not a real-time assessment based on, for instance, location or real supermarket supplies) but another constant assigned to a product. According to Points (4) and (5) in Section~\ref{subsubsec:Opacity2}, these should be explained to the user, particularly because considering Yuka does not give recommendation above score 50. (Therefore, the user cannot even figure out which product would be the 'healthiest' according to Yuka's classification.)
    \item \textbf{Regarding Fairness}:
    Yuka fails to deliver in all aspects proposed in Section~\ref{subsec:Fairness}: (1) As mentioned above, the lack of transparency in the recommending process does not provide clarity for the manufacturer and the database, the very design of the application (the lack of multi-stakeholder approach, in particular, the involvement of the manufacturers) practically makes it impossible to approximate verified, updated and complete data sets. (2) The principle of proportional fairness suffers every time when the ranking of the recommendations and scores do not correlate and (3) it is dubious, that the intentional 'product availability' variable is a feature in its current form, and not an unwanted bias.
    \item \textbf{Disrupting the Food Industry} This is the biggest worry of the authors, that with the arbitrary, not controlled, not scientifically validated interpretations of 'healthy' combined with the millions of users would push the manufacturers to adhere to distorted standards (c.f. Section~\ref{subsubsec:Disruptive}). Furthermore, once powerful stakeholders inspired by the marketing value of high Yuka scores, their lobbying power could further influence the society and even those who fail to see the benefit of this application pragmatically will be exposed to these products.  
    
\end{itemize}
\section{Conclusion}
\label{sec:Conclusion}

In this paper, we aimed to provide a bridge over the gap between the emerging ethical principles for beneficial AI and an existing domain of AI applications by narrowing down their interpretation by specifying the context. We proposed the useful distinction between F-RS(P-) and F-RS(P+), which helped us define the relevant and empirically meaningful ethical desiderata for the former in the spirit that it will be understandable from both the angles of regulators and designers. In the case of F-RS(P-) these ethical desiderata are revolved around three issues: first, we argued what sort of information needs to be provided regarding the business model, the ethical agenda, and the other variables affecting the recommendation, both in terms of its motivations and its consequences of the ordering; second, how a multi-stakeholder approach to the architecture and its implementation is necessary to maintain the beneficial nature of potentially disruptive effects on our food industry, to prevent a "race to the bottom" attitude regarding the user-friendliness of the required explanations; and third, the requirements to aim for updated, complete, validated and verified data sets that are robust enough to grant procedural fairness to the producers. This paper, however, has its limitations, which points to further research: First, creating an implementable multi-stakeholder architecture for F-RS(P-) that complies with these ethical desiderata. Second, how the emerged ethical desiderata can be implemented to our existing legal frameworks, in other words, how these ethical values could be translated to \emph{ex-ante} regulations and how \emph{ex-post} adjudications could cluster certain type of cases, and what legal tools are applicable to "fine-tune" the differences among the cases clustered together. Furthermore, since it is just a matter of time that personalised food-recommender systems become prevalent in the European market, how the ethical desiderata can be articulated for these cases in order to comply with the principles should be investigated shortly.

\bibliographystyle{ACM-Reference-Format}
\balance
\bibliography{references}

\end{document}